\documentclass[prl,twocolumn,floatfix,showpacs ]{revtex4-1}
\UseRawInputEncoding
\usepackage{amsmath}
\usepackage{amssymb}
\usepackage{graphicx}
\usepackage{dcolumn}
\usepackage{bm}
\usepackage[colorlinks=true,linkcolor=blue,anchorcolor=blue, citecolor=cyan,urlcolor=cyan]{hyperref}
\usepackage[mathlines]{lineno}
\usepackage{ulem}
\usepackage{epstopdf}
\usepackage{tabularx} 
\begin{document}
\title{Robust $d$-wave altermagnetism in  $\mathrm{XCr_2Y_2O}$  (X=K, Rb, Cs; Y=S, Se, Te) family}
\author{San-Dong Guo}
\email{sandongyuwang@163.com}
\affiliation{School of Electronic Engineering, Xi'an University of Posts and Telecommunications, Xi'an 710121, China}

\begin{abstract}
 The $\mathrm{KV_2Se_2O}$,  $\mathrm{Rb_{1-\delta}V_2Te_2O}$  and $\mathrm{Cs_{1-\delta}V_2Te_2O}$ are experimentally confirmed to adopt either C-type or G-type antiferromagnetic configuration, corresponding to apparent or hidden altermagnetism.  However, their nearly degenerate energies lead to inconsistent experimental assignments between the two antiferromagnetic configurations. Here, we predict that the experimentally synthesized  $\mathrm{RbCr_2Se_2O}$ is a robust $d$-wave altermagnetic metal, since the energy difference between C-type and G-type configurations is large, which is  independent of electron correlation strength and van der Waals interaction. Upon applying in-plane uniaxial strain, $\mathrm{RbCr_2Se_2O}$ can generate a net total magnetic moment via a direct piezomagnetic effect, which is distinct from semiconductor that typically requires carrier doping in addition to strain.
This provides an experimental strategy for distinguishing the G-type antiferromagnetic configuration, in which the total magnetic moment remains zero under uniaxial strain.  Our work presents an isostructural $d$-wave altermagnetic $\mathrm{RbCr_2Se_2O}$  analogous to $\mathrm{KV_2Se_2O}$,  $\mathrm{Rb_{1-\delta}V_2Te_2O}$  and $\mathrm{Cs_{1-\delta}V_2Te_2O}$, which can facilitate further experimental verification.
Furthermore, these results are universal across materials of this family $\mathrm{XCr_2Y_2O}$ (X=K, Rb, Cs; Y=S, Se, Te), thus expanding the family of altermagnets.

\end{abstract}
\maketitle
\textcolor[rgb]{0.00,0.00,1.00}{\textbf{Introduction.---}}
Altermagnetism represents a new class of collinear magnetism beyond conventional ferromagnets and antiferromagnets. It features fully compensated magnetic moments under  special crystal symmetry, while hosting spin-split electronic bands even in the absence of spin-orbit coupling\cite{k4,k5}. This unique combination gives rise to spin-polarized transport and anomalous Hall effects, making altermagnets promising for next-generation spintronic applications.
 A variety of altermagnetic (AM) materials have been successively predicted theoretically and identified experimentally in recent years\cite{k4,k5,k5-1,k6,k6-1,k6-2,k6-3,k6-311, ex0,ex01,ex02,ex1,ex2,ex3,ex4,ex5,ex51,ex52,ex53}, which has greatly promoted the rapid development of altermagnetism as a vibrant research direction in condensed matter physics and spintronics.

Two-dimensional (2D) $d$-wave altermagnets exhibit distinct advantages in generating and manipulating spin currents, rendering them highly appealing for spintronic applications.  The $\mathrm{KV_2Se_2O}$,  $\mathrm{Rb_{1-\delta}V_2Te_2O}$  and $\mathrm{Cs_{1-\delta}V_2Te_2O}$  have been experimentally synthesized and can be regarded as quasi-two-dimensional materials\cite{ex3,ex4,ex5,ex51,ex52,ex53}. Such layered structures can give rise to apparent and hidden electronic states depending on different magnetic configurations\cite{h2,h3,h4,h5,h6,h7}. These materials possess two lowest-energy antiferromagnetic (AFM) configurations, namely the C-type (intralayer AFM, interlayer ferromagnetic (FM)) and G-type (both intralayer and interlayer AFM).
The C-type corresponds to apparent  altermagnetism, whereas the G-type corresponds to hidden altermagnetism (While no spin splitting is detected at the global level, local AM spin splitting is still  present.), as has been clearly proposed\cite{h6}.
Recently, hidden altermagnetism has also attracted increasing research attention. Hidden AM spin splitting has been predicted in multiferroic collinear AFM  $\mathrm{MnS_2}$, giving rise to various emergent responses\cite{ha1}, and tunable hidden AM splitting has also been reported in layered Ruddlesden-Popper oxides\cite{ha2}. Moreover, hidden altermagnetism is predicted to exist in
$\mathrm{Sr_{n+1}Cr_n O_{3n+1}}$  owing to orbital ordering rather than lattice symmetry\cite{ha2-1}.

Experimentally, the $\mathrm{KV_2Se_2O}$ and $\mathrm{Rb_{1-\delta}V_2Te_2O}$ have been identified to adopt the C-type AFM configuration\cite{ex3,ex4}, corresponding to apparent  altermagnetism, while the $\mathrm{Cs_{1-\delta}V_2Te_2O}$ exhibits the G-type configuration corresponding to hidden altermagnetism\cite{ex5}. However, for $\mathrm{KV_2Se_2O}$  and $\mathrm{Rb_{1-\delta}V_2Te_2O}$, two experiments yield contradictory results: one assigns them to the C-type\cite{ex3,ex4}, whereas the other determines them as the G-type\cite{ex52,ex53}. For this family of materials, the C-type and G-type configurations are nearly degenerate with a tiny energy difference\cite{edv}, so the vacancy distribution of intercalated atoms (K, Rb or Cs) can significantly influence the magnetic structure experimentally\cite{ex5,ex51}. Given that angle-resolved photoemission spectroscopy (ARPES) is surface-sensitive and its measurements of altermagnetism are prone to domain effects, differentiating apparent from hidden altermagnetism is not straightforward using only ARPES spectra. Recently, we have proposed that  the in-plane uniaxial strain can be used to  distinguish the C-type and G-type\cite{gsd}. For the C-type phase, uniaxial strain can induce a net magnetic moment, while the total moment of the G-type phase remains zero.

\begin{figure*}[t]
    \centering
    \includegraphics[width=0.90\textwidth]{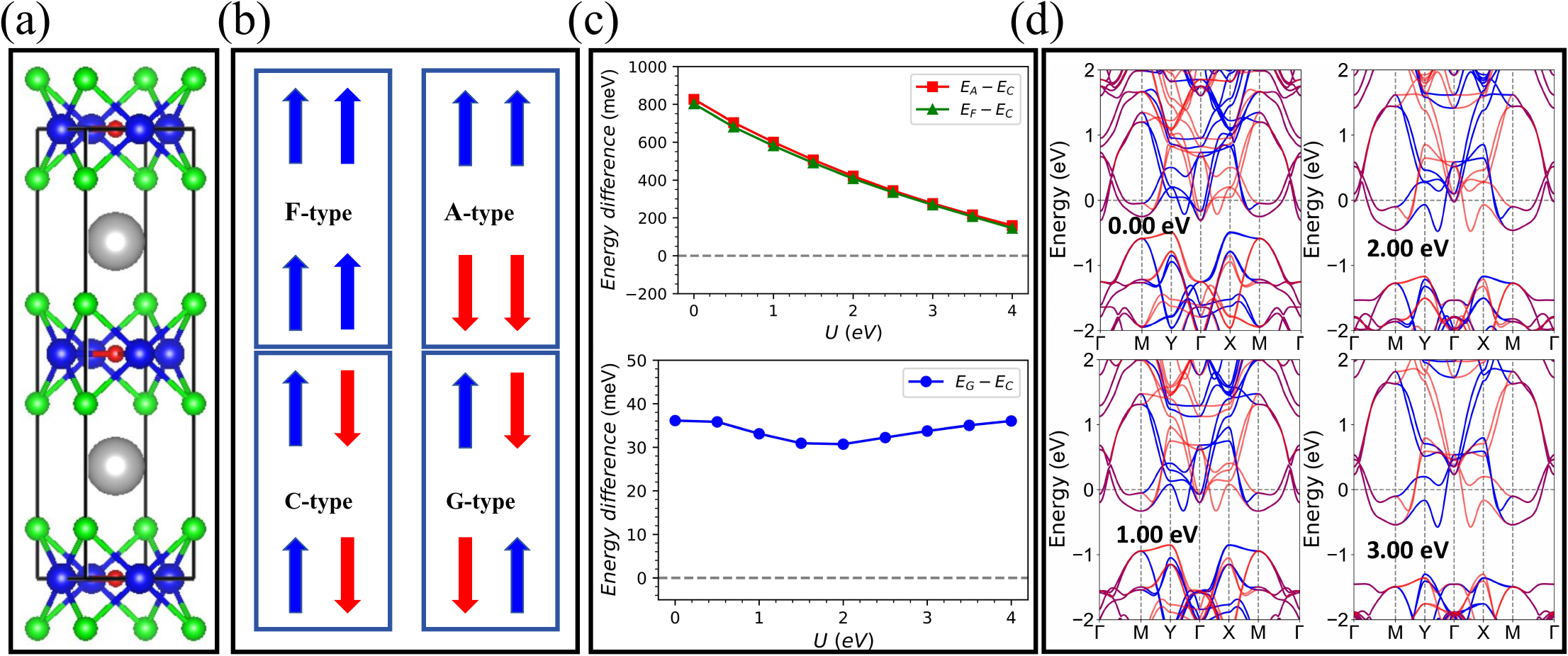}
    \caption{(Color online) For $\mathrm{RbCr_2Se_2O}$, (a): the crystal structure with blue, red, green and gray spheres representing Cr, O, Se and Rb atoms, respectively. The black  box denotes the magnetic primitive cell. (b): four
possible  magnetic  configurations  with F-type, A-type,  C-type  and G-type. (c): the energies (per magnetic primitive cell) of F-, A-, and G-type configurations as functions of $U$, with C-type set to zero. (d):  the global  energy  band structure with C-type AFM configuration at $U$=0.00, 1.00, 2.00 and 3.00 eV. The blue, red, and purple curves denote the spin-up, spin-down, and spin-degenerate bands, respectively. }\label{a}
\end{figure*}

\begin{figure}[t]
    \centering
    \includegraphics[width=0.45\textwidth]{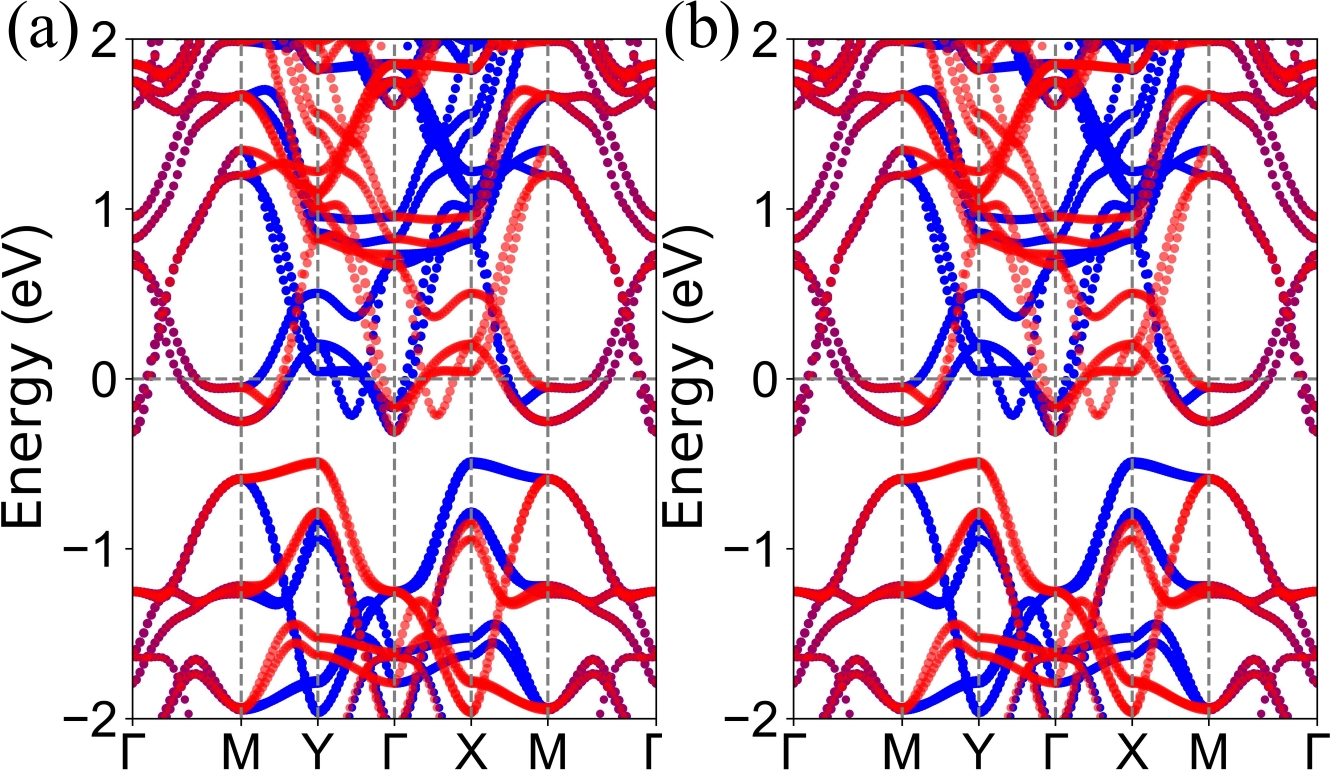}
    \caption{(Color online)  For $\mathrm{RbCr_2Se_2O}$ without uniaxial strain at $U$=0.00 eV,   the energy band structures with spin-resolved projections onto the sector A (a) and sector B (b) with C-type AFM configuration. The blue, red, and purple denote the spin-up, spin-down, and spin-degenerate bands, and the weighting coefficient is proportional to the circle size.}\label{b}
\end{figure}

It is natural to ask whether other isostructural compounds can possess robust apparent  altermagnetism. These materials are layered structures constructed by stacking 2D AM materials $\mathrm{V_2Se_2O}$ or  $\mathrm{V_2Te_2O}$, with K, Rb, or Cs atoms intercalated.
Recently, a large number of 2D AM materials isostructural to  $\mathrm{V_2Se_2O}$ or  $\mathrm{V_2Te_2O}$ have been predicted\cite{dc1}, providing a foundation for constructing such layered materials.  Fortunately, a new isostructural chromium oxyselenide, $\mathrm{RbCr_2Se_2O}$, has recently been synthesized experimentally\cite{dc2}, which can be regarded as being constructed by intercalating Rb atoms into $\mathrm{Cr_2Se_2O}$ bilayers.  Magnetic susceptibility measurements indicate that $\mathrm{RbCr_2Se_2O}$ undergoes an AFM transition at 345 K. Here, using first-principles calculations, we establish that $\mathrm{RbCr_2Se_2O}$ is a robust $d$-wave AM metal, and uniaxial strain can induce an experimentally observable net magnetic moment. These results are universal across the family of $\mathrm{XCr_2Y_2O}$  materials (X=K, Rb, Cs; Y=S, Se, Te).
These Cr-based systems are all AM metals, which show unique advantages in exploring physical phenomena related to low-energy quasiparticle excitations and enabling spintronic applications. Thanks to the finite electrical conductivity of metals, spin currents can be directly manipulated by electric fields.

\textcolor[rgb]{0.00,0.00,1.00}{\textbf{Computational detail.---}}
We perform density functional theory (DFT) calculations\cite{1,111} using the Vienna ab initio simulation package (VASP)\cite{pv1,pv2,pv3} within the framework of the projector augmented-wave (PAW) method. The generalized gradient approximation (GGA) of  Perdew, Burke, and Ernzerhof (PBE)\cite{pbe} is adopted as the exchange-correlation functional.
A kinetic energy cutoff of 500 eV, a total energy convergence criterion of $10^{-8}$ eV, and a force convergence criterion of 0.001$\mathrm{eV\cdot{\AA}^{-1}}$ are used to confirm reliable results. A 14$\times$14$\times$2 Monkhorst-Pack $k$-point meshes are  employed to sample the Brillouin zone (BZ) for both structural relaxation and electronic structure calculations.  To examine the robustness of the energy difference between magnetic configurations, the Hubbard correction is incorporated within the rotationally invariant approach proposed by Dudarev et al.\cite{du},  and  the van der Waals (vdW) interaction with the dispersion-corrected DFT-D3 method\cite{dft3} is also considered. When uniaxial strain is applied along the $a$-axis, both the $b$- and $c$-axis lattice parameters are fully relaxed.

\begin{figure*}[t]
    \centering
    \includegraphics[width=0.90\textwidth]{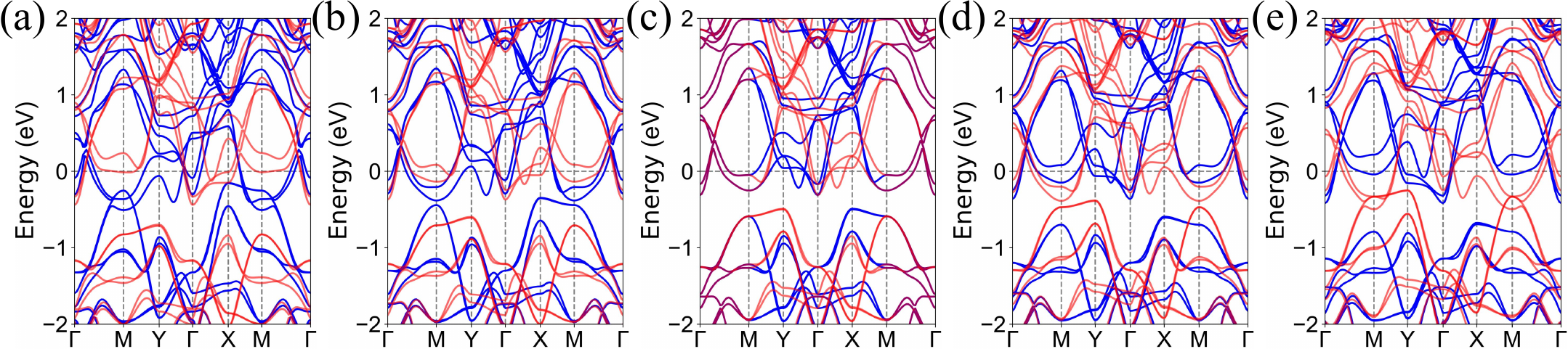}
     \caption{(Color online)  For $\mathrm{RbCr_2Se_2O}$,   the global  energy  band structure with C-type AFM configuration at $a/a_0$=0.96 (a), 0.98 (b), 1.00 (c), 1.02 (d) and 1.04 (e). The blue, red, and purple curves denote the spin-up, spin-down, and spin-degenerate bands, respectively. }\label{c}
\end{figure*}

\begin{figure}[t]
    \centering
    \includegraphics[width=0.40\textwidth]{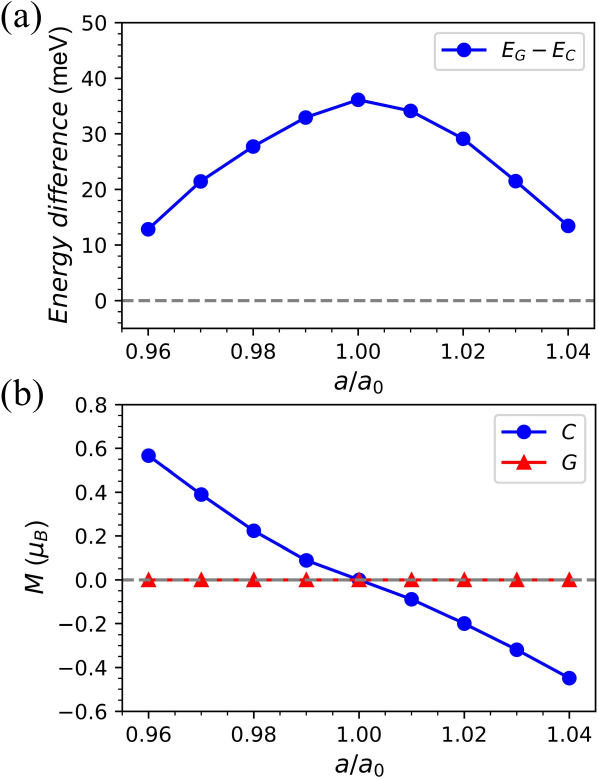}
     \caption{(Color online)  For    $\mathrm{RbCr_2Se_2O}$, (a):  the energy (per magnetic primitive cell) of  G-type configuration as function of $a/a_0$ with C-type set to zero.   (b): the total magnetic moment as a function of $a/a_0$ with C-type  and G-type AFM configurations.}\label{d}
\end{figure}
\begin{figure*}[t]
    \centering
    \includegraphics[width=0.8\textwidth]{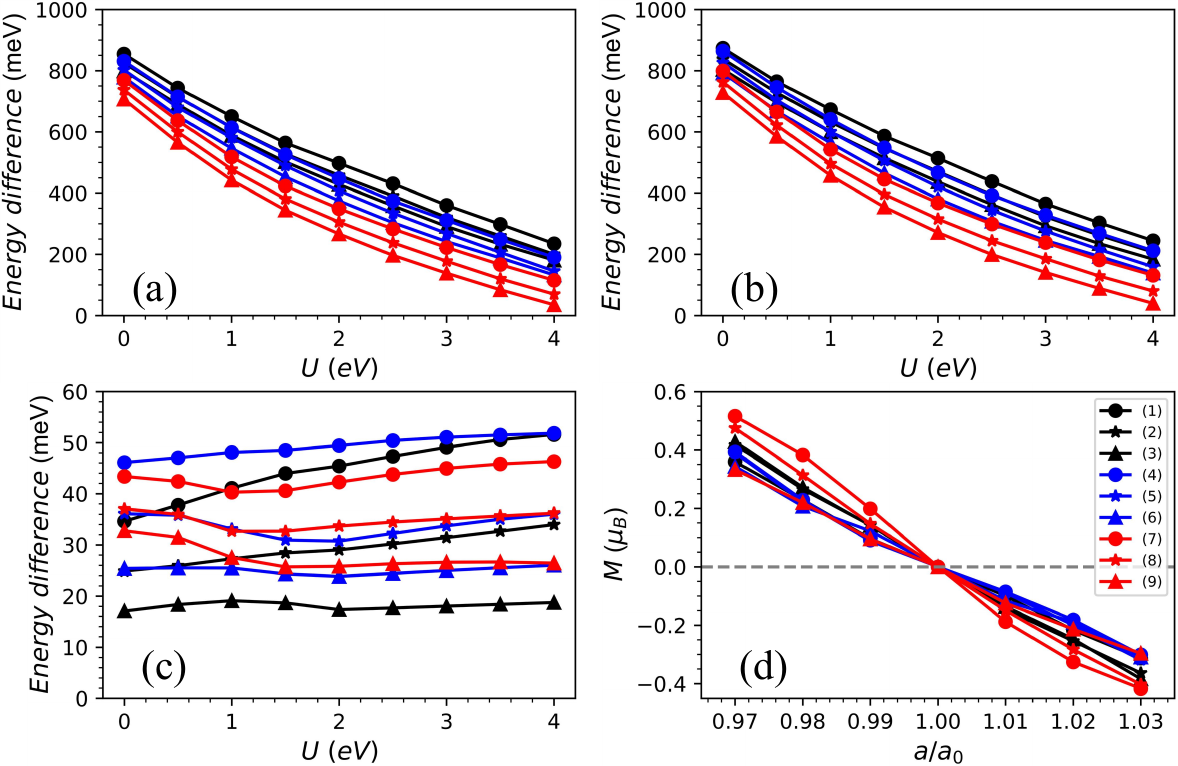}
     \caption{(Color online)  For $\mathrm{KCr_2S_2O}$ (1),   $\mathrm{RbCr_2S_2O}$ (2),   $\mathrm{CsCr_2S_2O}$ (3),  $\mathrm{KCr_2Se_2O}$ (4),   $\mathrm{RbCr_2Se_2O}$ (5),   $\mathrm{CsCr_2Se_2O}$ (6),   $\mathrm{KCr_2Te_2O}$ (7),   $\mathrm{RbCr_2Te_2O}$ (8) and     $\mathrm{CsCr_2Te_2O}$ (9),   the energies (per magnetic primitive cell) of F-type (a), A-type (b) and  G-type (c)  configurations as a function of $U$ with C-type set to zero, and  the total magnetic moment (d)  as a function of $a/a_0$ with C-type. }\label{e}
\end{figure*}

\textcolor[rgb]{0.00,0.00,1.00}{\textbf{Main results.---}}
As shown in \autoref{a} (a),  the experimentally synthesized $\mathrm{RbCr_2Se_2O}$ is a quasi-two-dimensional compound composed of alternating Rb and $\mathrm{Cr_2Se_2O}$ layers, crystallizing in the $P4/mmm$ space group (No.123)\cite{dc2}, which  shares the same crystal structure as the  $\mathrm{KV_2Se_2O}$,  $\mathrm{Rb_{1-\delta}V_2Te_2O}$  and $\mathrm{Cs_{1-\delta}V_2Te_2O}$\cite{ex3,ex4,ex5,ex51,ex52}. This structure is also consistent with our originally proposed model of hidden altermagnetism\cite{h6}. And then, we define the top $\mathrm{Cr_2Se_2O}$ layer as sector A and the bottom $\mathrm{Cr_2Se_2O}$ layer as sector B.
Based on   \autoref{a} (b), four possible magnetic configurations are considered, namely FM intralayer with FM interlayer coupling, FM intralayer with AFM interlayer coupling, AFM intralayer with FM interlayer coupling, and AFM intralayer with AFM interlayer coupling, which are defined as F-type, A-type, C-type, and G-type, respectively. The magnetic configuration significantly affects both the spin-space group symmetry $[C_2||O]$ (The $C_2$ denotes a twofold rotation in spin space, while the  $O$ represents  mirror ($M$),   rotation ($C$), etc in lattice space.) and the corresponding spin splittings in momentum space\cite{ex5,ex51}. The A-type, C-type, and G-type configurations are all AFM, but exhibit distinct symmetry and magnetic characters.
The A-type structure respects the $[C_2||M_z]$ symmetry, with globally spin-degenerate band structures but local FM spin splitting, which can be described as hidden ferromagnetism.  The C-type structure obeys the $[C_2||C_4]$ symmetry, displaying $d$-wave altermagnetism in its band structure.
The G-type structure preserves $PT$ (the joint symmetry ($PT$) of space inversion symmetry ($P$) and time-reversal symmetry ($T$)) symmetry, featuring globally spin-degenerate bands but local AM spin splitting, referred to as hidden altermagnetism\cite{h6}.

We next determine the ground-state magnetic configuration of $\mathrm{RbCr_2Se_2O}$. Taking the C-type magnetic configuration as the reference, the energies of the G-type, A-type, and F-type configurations as a function of $U$ are plotted in \autoref{a} (c).
It can be clearly seen that, within the considered range of $U$, the energies of the A-type and F-type configurations are much higher than that of the C-type configuration. Although the energies of the G-type and C-type configurations are close to each other, their energy difference is significantly larger than those between the two magnetic configurations in  $\mathrm{XV_2Y_2O}$ (X=K, Rb, Cs; Y=S, Se, Te) systems (Less than 4.5 meV per magnetic primitive cell\cite{ex5,ex51,edv}), which is more favorable for the unambiguous experimental determination of the ground state of $\mathrm{RbCr_2Se_2O}$.
To verify the reliability of our results, we also considered the vdW interaction. The energies of the four magnetic configurations are presented in FIG.S1\cite{bc}. The vdW interaction does not affect the essential conclusions, and  the energy difference between the G-type and C-type configurations still remains large.

The global band structures of the  C-type magnetic configuration at typical $U$ values of 0.00, 1.00, 2.00 and 3.00 eV are plotted in \autoref{a} (d).
It is clearly shown that $\mathrm{RbCr_2Se_2O}$ exhibits $d$-wave altermagnetism governed by $[C_2||C_4]$ symmetry, with spin degeneracy along the $\Gamma$-M path and alternating spin splitting across the M-Y-$\Gamma$ and M-X-$\Gamma$ paths.
Compared with $\mathrm{KV_2Se_2O}$,  $\mathrm{Rb_{1-\delta}V_2Te_2O}$  and $\mathrm{Cs_{1-\delta}V_2Te_2O}$\cite{ex3,ex4,ex5,ex51}, the most distinct difference is that an obvious energy gap exists below the Fermi level in $\mathrm{RbCr_2Se_2O}$. As $U$ increases, the bands near the Fermi level become sparse.
Since the band structure of $\mathrm{Cs_{1-\delta}V_2Te_2O}$ and other systems at $U$=0.00 eV is more consistent with experimental results\cite{ex5,ex51}, we mainly focus on the case of $U$=0.00 eV for $\mathrm{RbCr_2Se_2O}$ in the following discussion, and the essential conclusions are independent of $U$.
For $\mathrm{RbCr_2Se_2O}$  without uniaxial strain at $U$=0.00 eV, the spin-resolved band structures projected onto sector A and sector B for the C-type AFM configuration are plotted in \autoref {b}. The projected band structures  show that the two sectors A and B are fully equivalent. If each sector carries a net magnetic moment, their moments should be equal in magnitude and constructively additive. This differs from the G-type case, where the two sectors exhibit opposite spin polarizations, and any net magnetic moments thus cancel each other out\cite{gsd}.
The calculated absolute value of the magnetic moment on the Cr atom is 2.89  $\mathrm{\mu_B}$, which is  close to that expected for  $\mathrm{Cr^{3+}}$
(3 $\mathrm{\mu_B}$). The magnetization direction can be determined by the magnetic anisotropy energy (MAE), which is defined as  energy difference $E_x$-$E_z$, where $E_x$/$E_z$ is the energy per magnetic primitive cell  when the magnetization is along the $x$/$z$ direction.
The calculated MAE is 210 $\mathrm{\mu eV}$, and the positive value indicates that material  $\mathrm{RbCr_2Se_2O}$ exhibits an out-of-plane magnetization direction.

In our previous work, we proposed that uniaxial strain can induce magnetic moments in C-type $\mathrm{KV_2Se_2O}$,  $\mathrm{Rb_{1-\delta}V_2Te_2O}$  and $\mathrm{Cs_{1-\delta}V_2Te_2O}$ systems, whereas the magnetic moment remains zero in the G-type magnetic configuration\cite{gsd}. This provides an experimentally feasible strategy to distinguish between C-type and G-type. Here, we also investigate the effects of $a$-axis uniaxial strain on electronic structures and magnetic properties of  $\mathrm{RbCr_2Se_2O}$ using the strain parameter
$a/a_0$ (0.95-1.05), where $a$ and $a_0$ denote the strained and equilibrium lattice constants, respectively.
The global  band structures of
 $\mathrm{RbCr_2Se_2O}$
 with C-type at $a/a_0$=0.96, 0.98, 1.00, 1.02 and 1.04 are shown in \autoref{c}, and those with G-type  in FIG.S2\cite{bc}.
 The energy of  G-type configuration  and  the total magnetic moment with C-type  and G-type as a function of $a/a_0$ are plotted in \autoref{d}.

 According to \autoref{d} (a),    the C-type configuration of  strained $\mathrm{RbCr_2Se_2O}$  remains the ground state at all times.
Under uniaxial strain, the C-type configuration exhibits spin splitting across the entire BZ, corresponding to the so-called
$s$-wave symmetry. Combined with a nonzero total magnetic moment (see \autoref{d} (b)), this corresponds to the transition from altermagnetism to ferrimagnetism induced by uniaxial strain\cite{gsd}.  As the strain changes from compression to tension, the total magnetic moment varies nearly linearly from positive to negative values. At
$a/a_0$=0.97, the magnetic moment reaches 0.39 $\mu_B$, which is readily measurable in experiments.
Uniaxial strain induces asymmetry in the band structures along the M-Y-$\Gamma$ and M-X-$\Gamma$ paths, and this asymmetry is reversed when the strain switches from compression to tension. For the G-type configuration, the system retains $PT$-antiferromagnetism with spin-degenerate bands and  a vanishing total magnetic moment, while the uniaxial-strain-induced band asymmetry still persists.
At the local scale, uniaxial strain induces a transition from hidden altermagnetism  to  hidden ferrimagnetism\cite{gsd}.
We also consider the case of $U$=3.00 eV, and all essential results remain unchanged at least under small strain (see FIG.S3 and FIGS4\cite{bc}).
 Uniaxial strain indeed can also be employed in $\mathrm{RbCr_2Se_2O}$ to distinguish between C-type and G-type magnetic configurations, corresponding to apparent and hidden altermagnetism, respectively.

\begin{figure}[t]
    \centering
    \includegraphics[width=0.45\textwidth]{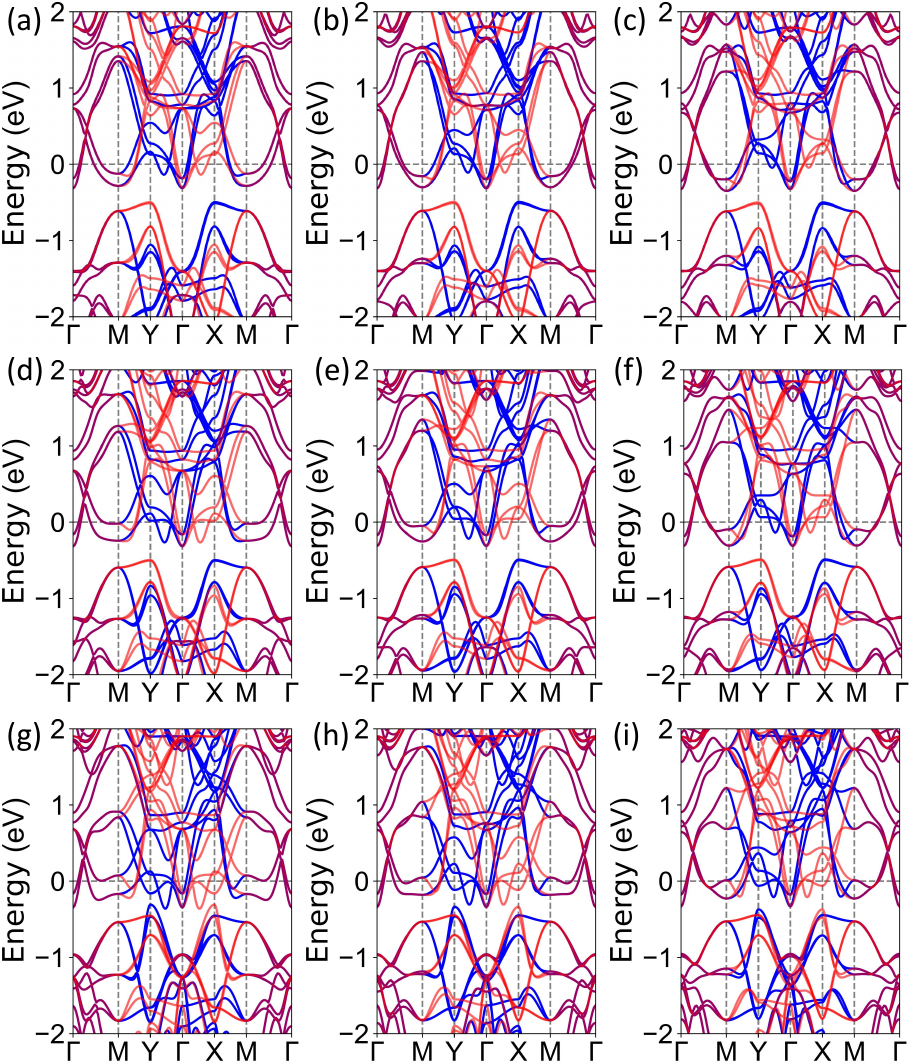}
     \caption{(Color online)  For $\mathrm{KCr_2S_2O}$ (a),   $\mathrm{RbCr_2S_2O}$ (b),   $\mathrm{CsCr_2S_2O}$ (c),  $\mathrm{KCr_2Se_2O}$ (d),   $\mathrm{RbCr_2Se_2O}$ (e),   $\mathrm{CsCr_2Se_2O}$ (f),   $\mathrm{KCr_2Te_2O}$ (g),   $\mathrm{RbCr_2Te_2O}$ (h) and     $\mathrm{CsCr_2Te_2O}$ (i),  the global  energy  band structure with C-type by using  GGA, and the blue, red, and purple curves denote the spin-up, spin-down, and spin-degenerate bands, respectively.  }\label{f}
\end{figure}

\textcolor[rgb]{0.00,0.00,1.00}{\textbf{Discussion and Conclusion.---}}
We also investigate  the electronic structures and magnetic properties of  the entire $\mathrm{XCr_2Y_2O}$ (X=K, Rb, Cs; Y=S, Se, Te) family to verify the generality of the above results.  First, the dependence of the lattice parameters
$a$ and $c$  on X and Y are presented in FIG.S5\cite{bc}.
It is found that $a$ shows little variation with X and Y, whereas
$c$ is mainly determined by Y and then by X. Specifically, all compounds with
Y=Te  have larger $c$ than those with Y=S or Se, and $c$ increases gradually as X changes from K to Rb to Cs.
This trend is also consistent with that observed in $\mathrm{KV_2Se_2O}$,  $\mathrm{Rb_{1-\delta}V_2Te_2O}$  and $\mathrm{Cs_{1-\delta}V_2Te_2O}$\cite{ex51}.
For these nine compounds, with the C-type configuration as the reference, the energies of the F-type, A-type and  G-type phases as a function of
$U$ are plotted in \autoref{e} (a, b, c).
In all cases, the energies of F-type and A-type are considerably higher than that of C-type, and the  G-type exhibits a distinct energy difference from C-type.
The energy difference between the G-type and C-type magnetic configurations decreases gradually from K to Rb to Cs,  when the chalcogen atom (S, Se, Te) is fixed. This phenomenon can be attributed to the gradual increase of lattice parameter $c$ (see  FIG.S5\cite{bc}), which weakens the interlayer interaction and thereby facilitates the reversal of the $\mathrm{N\acute{e}el}$ vector in a single layer $\mathrm{Cr_2Y_2O}$.
The global band structures of the nine compounds with C-type ground-state magnetic configuration are displayed in \autoref{f} by using GGA,  and those  calculated using the GGA+$U$  ($U$=3.00 eV) are also plotted in  FIG.S6\cite{bc}.  For all cases, they exhibit highly similar band features, showing  $d$-wave spin-splitting symmetry. Compared with V-based materials, Cr-based compounds exhibit highly analogous band structures. The primary distinction lies in that each magnetic unit cell of Cr-based systems contains four additional electrons, which raises the Fermi level upward within the conduction band.
In all cases, the total magnetic moments as a function of uniaxial strain $a/a_0$ are plotted in \autoref{e} (d).
The total magnetic moments of   all systems exhibit identical strain dependence, and  uniaxial strain can induce a sizable magnetic moment in every case, which is favorable for experimental determination.
\begin{figure}[t]
    \centering
    \includegraphics[width=0.45\textwidth]{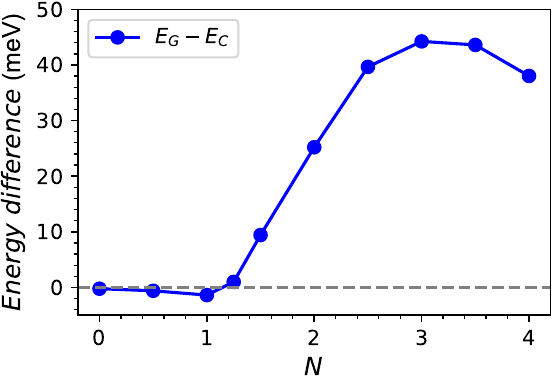}
     \caption{(Color online)  For    $\mathrm{CsV_2Te_2O}$,  the energy (per magnetic primitive cell) of  G-type configuration as a function of  number of doped electrons ($N$)  with C-type set to zero.}\label{g}
\end{figure}
\begin{figure}[t]
    \centering
    \includegraphics[width=0.45\textwidth]{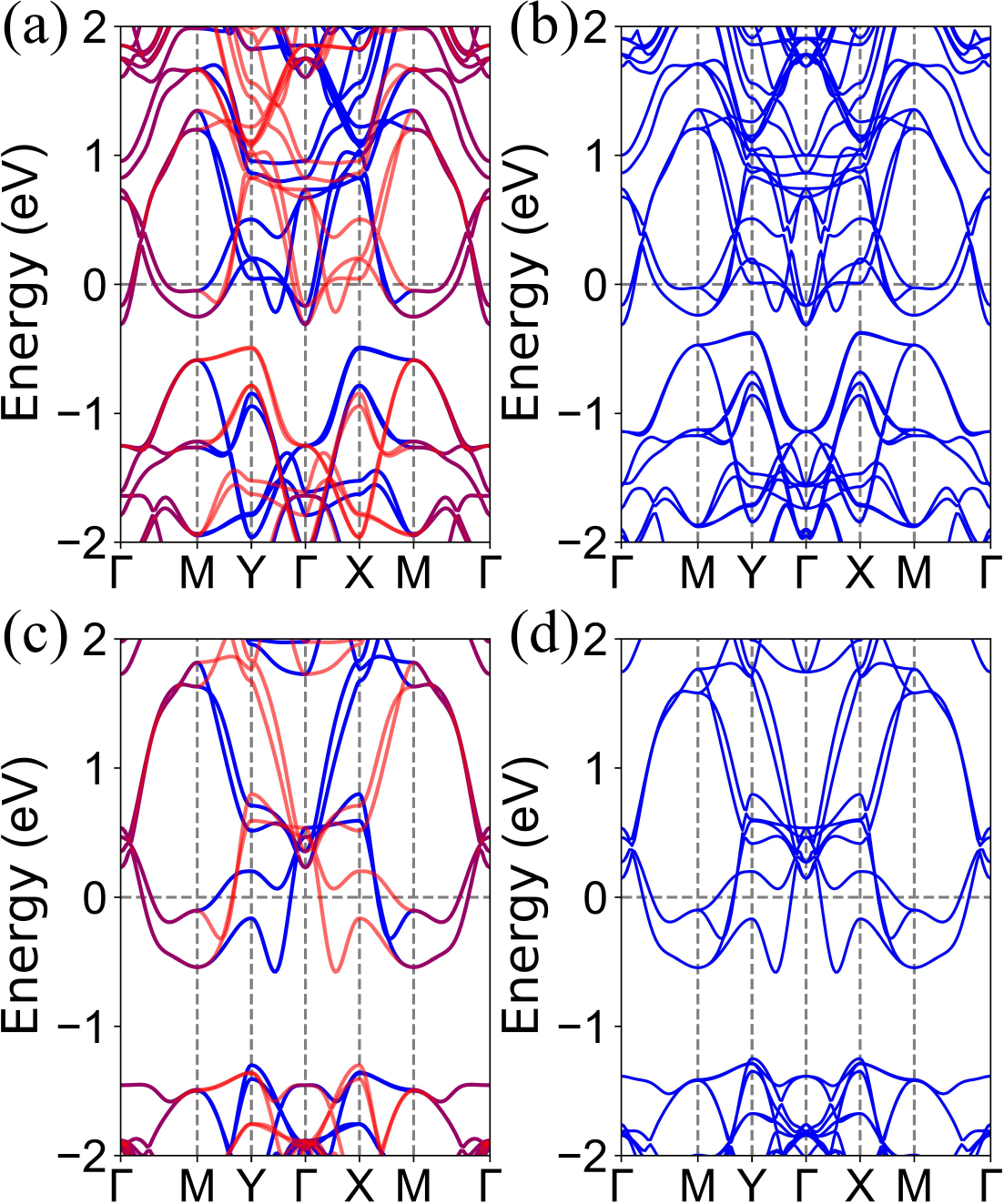}
     \caption{(Color online)  For   $\mathrm{RbCr_2Se_2O}$,  the energy band structures by using GGA (a), GGA+SOC (b), GGA+$U$ ($U$=3.00 eV) and GGA+SOC+$U$ ($U$=3.00 eV) (d). In (a, c), the blue, red, and purple curves denote the spin-up, spin-down, and spin-degenerate bands, respectively. }\label{h}
\end{figure}

Experimentally, compound $\mathrm{RbCr_2Se_2O}$ is identified as a semiconductor, which contradicts our theoretical predictions. This discrepancy is most likely attributed to the fact that the experimentally synthesized $\mathrm{RbCr_2Se_2O}$ is polycrystalline rather than single-crystalline. Nevertheless, another recently synthesized sister compound $\mathrm{CsCr_2S_2O}$\cite{expp} has been experimentally confirmed to be metallic when adopting the high-temperature crystal structure proposed by us, which is in good agreement with our theoretical predictions.

The Cr-based system exhibits a more robust C-type magnetic configuration than the V-based system. An intuitive explanation is that Cr possesses one more valence electron than V; that is, the increase in valence electron number stabilizes the C-type magnetic configuration.
To verify our conjecture, we take the V-based material  $\mathrm{CsV_2Te_2O}$ as an example and calculate the energy difference between the G-type and C-type magnetic configurations as a function of doped electron number $N$. The results are plotted in \autoref{g}. For one magnetic unit cell, the Cr-based system has four more electrons than the V-based system; hence, $N$  is set to range from 0 to 4. It can be clearly seen that at low doped electron numbers, the energy difference between the C-type and G-type magnetic configurations is very small. When $N$ is around 3, the energy of the G-type configuration becomes distinctly higher than that of the C-type. For the Cr-based system, the energy difference on the order of tens of meV is sufficient to resist the effects of temperature, structural disorder and non-stoichiometry.  Experimentally, X (X=K, Rb, Cs)  vacancies are commonly present, which slightly reduces the electron count of the system. As shown in \autoref{g}, such vacancies hardly affect the energy difference between G-type and C-type magnetic configurations for Cr-based systems. By contrast, the intrinsic energy difference is rather small in V-based systems, making the ground-state magnetic configuration highly susceptible to vacancy effects. This accounts for the inconsistent magnetic configurations observed in different experiments\cite{ex3,ex4,ex52,ex53}.

Finally, we also investigate the effect of spin-orbit coupling (SOC) on the band structure of Cr-based family materials. Taking material $\mathrm{RbCr_2Se_2O}$ as an example, the corresponding band structures are presented in \autoref{h}. The calculated results reveal that SOC has a negligible influence on the electronic bands near the Fermi level.

In summary,  the experimentally synthesized
$\mathrm{RbCr_2Se_2O}$ is predicted to be  a robust
$d$-wave altermagnet, supported by a large energy difference between the C-type and G-type magnetic configurations. Under in-plane uniaxial strain,
the $\mathrm{RbCr_2Se_2O}$ with C-type  can generate a net total magnetic moment through a direct piezomagnetic effect, which  offers an experimental approach to distinguish the G-type AFM configuration with the total magnetic moment remaining zero even under uniaxial strain.
Our work facilitates the experimental verification and realization of robust $d$-wave AM  $\mathrm{XCr_2Y_2O}$ (X=K, Rb, Cs; Y=S, Se, Te) family.

~~~~\\
\textbf{Remark}: After we submitted the preprint (arXiv:2604.00412),  we have noted that compound  $\mathrm{CsCr_2S_2O}$ has also been successfully synthesized experimentally (see arXiv:2604.02114).

\begin{acknowledgments}
This work is supported by Natural Science Basis Research Plan in Shaanxi Province of China   (2025JC-YBMS-008). We are grateful to Shanxi Supercomputing Center of China, and the calculations were performed on TianHe-2. We thank Prof. Guangzhao Wang and Prof. Yang Liu for providing VASP software  and helpful discussions.
\end{acknowledgments}


\begin{references}

\bibitem{k4}L. $\mathrm{\breve{S}}$mejkal, J. Sinova and T. Jungwirth, Beyond conventional ferromagnetism and antiferromagnetism: A phase with nonrelativistic spin and crystal rotation symmetry,  Phys. Rev. X \textbf{12}, 031042 (2022).


\bibitem{k5}I. Mazin, Altermagnetism-a new punch line of fundamental magnetism,  Phys. Rev. X \textbf{12}, 040002 (2022).

\bibitem{k5-1}L. Bai, W. Feng, S. Liu, L. $\mathrm{\breve{S}}$mejkal, Y. Mokrousov, and Y. Yao, Altermagnetism: Exploring New Frontiers in Magnetism and Spintronics, Adv. Funct. Mater. \textbf{34},  2409327 (2024).

\bibitem{k6}H.-Y. Ma, M. L. Hu, N. N. Li, J. P. Liu, W.
Yao, J. F. Jia and J. W.  Liu, Multifunctional antiferromagnetic materials with giant piezomagnetism and noncollinear spin current, Nat. Commun. \textbf{12}, 2846 (2021).

\bibitem{k6-1}Y. Liu, J. Yu and C. C. Liu, Twisted Magnetic Van der Waals Bilayers: An Ideal Platform for Altermagnetism, Phys. Rev. Lett. \textbf{133}, 206702 (2024).

\bibitem{k6-2}X. Chen, D. Wang, L. Y. Li and B. Sanyal, Giant spin-splitting and tunable spin-momentum locked transport in room temperature collinear antiferromagnetic semimetallic CrO monolayer, Appl. Phys. Lett. \textbf{123}, 022402 (2023).

\bibitem{k6-3}B. Pan, P. Zhou, P. Lyu, H. Xiao, X. Yang, and L. Sun, General stacking theory for
altermagnetism in bilayer systems, Phys. Rev. Lett. \textbf{133}, 166701 (2024).

\bibitem{k6-311}L. Zhang and G. Gao, Dimension- and Facet-Dependent Altermagnetic Biferroics and Ferromagnetic Biferroics and Triferroics in CrSb, Adv. Funct. Mater.  \textbf{36}, e25978 (2026).


\bibitem{ex0}H. Bai, L. Han, X. Y. Feng, Y. J. Zhou, R. X. Su,
Q. Wang, L. Y. Liao, W. X. Zhu, X. Z. Chen, F. Pan, X. L. Fan, and C. Song, Observation of spin splitting
torque in a collinear antiferromagnet $\mathrm{RuO_2}$, Phys. Rev.
Lett. \textbf{128}, 197202 (2022).

\bibitem{ex01}S. Lee, S. Lee, S. Jung, J. Jung, D. Kim, Y. Lee,
B. Seok, J. Kim, B. G. Park, L. Smejkal, C. J. Kang,
and C. Kim, Broken Kramers degeneracy in altermagnetic MnTe, Phys. Rev. Lett. \textbf{132}, 036702 (2024).



\bibitem{ex02}G. Yang, Z. Li, S. Yang, J. Li, H. Zheng, W. Zhu,
Z. Pan, Y. Xu, S. Cao, W. Zhao, A. Jana, J. Zhang,
M. Ye, Y. Song, L. H. Hu, L. Yang, J. Fujii, I. Vobornik,
M. Shi, H. Yuan, Y. Zhang, Y. Xu, and Y. Liu, Three-dimensional mapping of the altermagnetic spin splitting
in CrSb, Nat Commun \textbf{16}, 1442 (2025).

\bibitem{ex1}Z. Zhou,  X.  Cheng, M.  Hu, R. Chu, H. Bai, L. Han, J. Liu, F. Pan and C. Song,  Manipulation of the altermagnetic order in CrSb via crystal symmetry, Nature \textbf{638}, 645 (2025).


\bibitem{ex2}J. Ding, Z. Jiang, X. Chen, Z. Tao, Z. Liu, T. Li, J. Liu, J. Sun and  J. Cheng, Large Band Splitting in $g$-Wave Altermagnet CrSb, Phys. Rev. Lett. \textbf{133}, 206401 (2024).


\bibitem{ex3}B. Jiang,  M.  Hu, J.  Bai, Z. Song, C. Mu, G. Qu, W. Li, W. Zhu, H. Pi, Z. Wei, Y. J. Sun, Y. Huang, X. Zheng, Y. Peng, L. He, S. Li, J. Luo, Z. Li, G. Chen, H. Li, H. Weng and  T. Qian, A metallic room-temperature d-wave altermagnet, Nat. Phys. \textbf{21}, 754 (2025).

\bibitem{ex4}F. Zhang X. Cheng, Z. Yin, C. Liu, L. Deng, Y. Qiao, Z. Shi, S. Zhang, J. Lin, Z. Liu, M. Ye, Y. Huang, X. Meng, C. Zhang, T. Okuda, K. Shimada, S. Cui, Y. Zhao, G.-H. Cao, S. Qiao, J. Liu and C. Chen, Crystal-symmetry-paired spin-valley locking in a layered room-temperature metallic altermagnet candidate, Nature Phys. \textbf{21}, 760 (2025).
\bibitem{ex5}G. Yang, R. Chen, C.  Liu et al., Observation of hidden altermagnetism in $\mathrm{Cs_{1-\delta}V_2Te_2O}$,  arXiv:2512.00972 (2025).

\bibitem{ex51}C.-C. Liu, J. Li, J.-Y. Liu, J.-Y. Lu, H.-X. Li, Y. Liu  and G.-H. Cao, Physical properties and first-principles calculations of an altermagnet candidate $\mathrm{Cs_{1-\delta}V_2Te_2O}$, Phys. Rev. B \textbf{112}, 224439  (2025).

\bibitem{ex52}Y. Sun, Y. Huang, J.  Cheng et al., Antiferromagnetic structure of  $\mathrm{KV_2Se_2O}$: A neutron diffraction study, Phys. Rev. B \textbf{112}, 184416 (2025).

\bibitem{ex53}W. Xie,  C. Liu,  F. Zhang et al., G-type antiferromagnetic structure in  $\mathrm{Rb_{1-\delta}V_2Te_2O}$, arXiv:2604.17365 (2026).


\bibitem{h2}J.-X. Xiong, X. Zhang, L.-D. Yuan and  A. Zunger, Matter with apparent and hidden spin physics, Matter  \textbf{9}, 102674 (2026).


\bibitem{h3}S. Guan, J. X. Xiong, Z. Wang, and J. W. Luo, Progress of hidden spin polarization in inversion-symmetric crystals, Sci. China-Phys. Mech. Astron. \textbf{65}, 237301 (2022).


\bibitem{h4}L. D. Yuan, X. Zhang, C. M. Acosta and A. Zunger, Uncovering spin-orbit coupling-independent hidden spin polarization of energy bands in antiferromagnets, Nat. Commun. \textbf{14}, 5301 (2023).

\bibitem{h5} S. D. Guo and P. Zhou, Hidden half-metallicity, arXiv:2601.07128 (2026).

\bibitem{h6}S. D. Guo, Hidden altermagnetism, Front. Phys.   \textbf{21},  025201 (2026); arXiv:2411.13795 (2024).

\bibitem{h7}S. D. Guo, Hidden fully-compensated ferrimagnetism,  Phys. Chem. Chem. Phys.  \textbf{28}, 2188 (2026).

\bibitem{ha1}J. Matsuda, H. Watanabe and R. Arita, Multiferroic Collinear Antiferromagnets with Hidden Altermagnetic Spin Splitting, Phys. Rev. Lett. \textbf{134}, 226703 (2025).

\bibitem{ha2}T. Zhang, L. Yuan, J. M. Rondinelli, H. A. Fertig and S. Zhang, Tunable Hidden Altermagnetic Spin Splitting in Layered Ruddlesden¨CPopper Oxides,  Nano Lett.  \textbf{26}, 2778 (2026).

\bibitem{ha2-1}Q. N. Meier, A. Carta, C. Ederer and A. Cano, Net and Compensated Altermagnetism from Staggered Orbital Order: Layer-Dependent Spin Splitting in $\mathrm{Sr_{n+1}Cr_n O_{3n+1}}$, Phys. Rev. Lett. \textbf{136}, 116705  (2026).



\bibitem{edv}B. Thapa,  P.-H. Chang,  K.  Belashchenko  and I. I. Mazin, Is altermagnetism in vanadium oxychalcogenides a lost cause?, arXiv:2602.18672 (2026).

\bibitem{gsd}S. D. Guo and Y. Liu, Distinguishing apparent and hidden altermagnetism via uniaxial strain in $\mathrm{CsV_2Te_2O}$-family, arXiv:2603.25136   (2026).


\bibitem{dc1}R. Xu, Y. Gao and J. Liu, Chemical design of monolayer altermagnets, Natl. Sci. Rev.  \textbf{13}, nwaf528 (2026).


\bibitem{dc2}X. Sun, P. Chen, X. Wen  and H. Chen, Synthesis, Structure, and Physical Properties of  $\mathrm{RbCr_2Se_2O}$, Crystals \textbf{16}, 56 (2026).




\bibitem{1}P. Hohenberg and W. Kohn, Inhomogeneous Electron Gas, Phys. Rev. \textbf{136},
B864 (1964).

\bibitem{111} W. Kohn and L. J. Sham, Self-Consistent Equations Including Exchange and Correlation Effects, Phys. Rev. \textbf{140},
A1133 (1965).

\bibitem{pv1} G. Kresse, Ab initio molecular dynamics for liquid metals, J. Non-Cryst. Solids \textbf{193}, 222 (1995).

\bibitem{pv2} G. Kresse and J. Furthm$\ddot{u}$ller, Efficiency of ab-initio total energy calculations for metals and semiconductors using a plane-wave basis set, Comput. Mater. Sci. 6, \textbf{15} (1996).

\bibitem{pv3} G. Kresse and D. Joubert, From ultrasoft pseudopotentials to the projector augmented-wave method, Phys. Rev. B \textbf{59}, 1758 (1999).
\bibitem{pbe}J. P. Perdew, K. Burke and M. Ernzerhof, Generalized gradient approximation made simple, Phys. Rev. Lett. \textbf{77}, 3865 (1996).


\bibitem{du}S. L. Dudarev, G. A. Botton, S. Y. Savrasov, C. J. Humphreys, and A. P. Sutton, Electron-energy-loss spectra and the structural stability of nickel oxide: An LSDA+U study, Phys. Rev. B \textbf{57}, 1505 (1998).

\bibitem{dft3} S. Grimme, S. Ehrlich and L. Goerigk, Effect of the damping function in dispersion corrected density functional theory,  J. Comput. Chem. \textbf{32}, 1456 (2011).





\bibitem{bc}See Supplemental Material at [] for the  associated energy differences of magnetic configurations, band structures and lattice parameters.

\bibitem{expp}Y. Liu, C.-C.  Xu, J.-K. Bao et al., Altermagnetism and Room-Temperature Metal-to-Insulator
Transition in $\mathrm{CsCr_2S_2O}$, arXiv:2604.02114 (2026).





\end{references}
\end{document}